# Logic for Electromagnetic Field Patterns

GUENNADI A. KOUZAEV
Department of Electronics and Telecommunications
Norwegian University of Science and Technology (NTNU)
O.S. Bragstads plass 2B, Gløshaugen, No-7491 Trondheim
NORWAY
guennadi.kouzaev@iet.ntnu.no    http://www.iet.ntnu.no/groups/radio/people/kouzaev/index.htm

*Abstract:* - It is assumed that the digital-like spatio-time brain activity might be caused by non-topological transformations of patterns in the cortex which is a linear, analog and active system. Such an effect can be modeled by topologically modulated spatio-time electromagnetic signals which theory is proposed in this paper. The logical operations are performed by passive components, and a theory of them is considered. Two gates of this sort are simulated. A short review on semiconductor hardware for this type of spatial digital processing and computing is given.

## 1 Introduction

The brain and mind phenomena are an intriguing problem of XXI century. Although, full their understanding is hardly ever imaginable now, the research provides interesting intermediate results for neurophysiology, medicine, computer science and engineering. For example, the achievements of last century allowed establishing the role of synapses and dendrite networks in the brain signal processing. It was found that the brain is a pattern "machine", and its mechanism can be described by linear processing of waveforms [1-5]. Non-linear effects are involved mostly only for the axon communications when transmitting strong and relatively high-speed signals is necessary.

The brain cortex consists of several layers. They are connected to each other by vertical neurons and multitude dendrites and synapses. Patterns corresponding to the outer world are formed on the first layer of neurons. During the processing the signals from different layers are mixed, and the long-term memory is formed by distributed neuronal "circuits" in the whole brain.

Mostly, the brain dynamics is with low-level, waveform electrochemical signals and analog, linear processing of spatio-temporal patterns. Unfortunately, this knowledge, confirmed experimentally by many neurophysiologists, does not explain the logical or digital-like processing of information by brain.

In this submission, it is supposed to pay attention on the ignored fact that a system linear regarding to signal amplitudes can be essentially nonlinear regarding to their other parameters. Especially, it is valid for spatially modulated signals which shapes are described by nonlinear geometrical equations. Even simple operation of summation of such signals can lead to step-wise transformations of pattern spatial topology [6]. Probably, this spatio-temporal non-linearity can play a role in the brain "digital activity", and it encourages further research on the discrete processing of patterned signals by different means.

The goal of this paper is not the development of a new speculative theory of neural processing but discovering some mechanisms of spatio-temporal signal processing by artificial circuits hoping that the derived results can be useful for the brain science.

It is supposed that a bit of a patterned signal is represented by spatial topology. Two shapes are topologically different if they cannot be transformed by bi-continuous homeomorphic transformations to each other. Logical units can be assigned to these topologically different shapes that it was proposed by McKensey and Tarsky [7]. According to the author best knowledge, the first topologically modulated electromagnetic signals and hardware were proposed and studied in [8,9]. Further readings on spatial logic are from [10].

In this paper, we consider the switching mechanism of topologically modulated electromagnetic signals using ideas on the manifold theory [11-14]. A couple gates of a passive design is considered here.

## 2 Mathematical Models of Topologically Modulated Electromagnetic Signals

The spatially modulated electromagnetic signals are described by their field–force lines, calculated at a certain moment of time *t*. Their equations are the dynamical systems, and the temporal parameter *t* is a bifurcation parameter:

$$\frac{d\mathbf{r}_{e,h}(t)}{ds_{e,h}} = \mathbf{E}, \mathbf{H}(\mathbf{r}_{e,h}, t) \qquad (1)$$

where $\mathbf{r}_{e,h}$ are the radius-vectors of the field-force lines of the electric $\mathbf{E}$ and magnetic $\mathbf{H}$ fields, accordingly, and $s$ is the parametrical variable. In general case, the equations (1) are the non-autonomous dynamical systems. Besides, the vector fields $\mathbf{E}(\mathbf{r},t)$ and $\mathbf{H}(\mathbf{r},t)$ are governed by Maxwell equations, boundary and initial conditions.

The above mentioned equations can be transformed into an autonomic dynamical system introducing a new variable $\tau$ with the goal to use the results of the well-developed theory of autonomous dynamical systems [15]:

$$\frac{d\mathbf{r}_{e,h}}{d\tau} = \frac{\mathbf{E},\mathbf{H}}{|\mathbf{E}|,|\mathbf{H}|}\left(\frac{ds_{e,h}}{d\tau}\right)^{-1},$$
$$\frac{d\tau}{dt}=1. \qquad (2)$$

Qualitatively, (1) and (2) are described by their 3-D or 4-D topological schemes ("skeletons"), correspondingly [14]. They are composed from separatrices of the field-force lines - vector manifolds. Additionally, (1) and their schemes have the equilibrium manifolds where the field is zero. The extended system (2) has no conventional equilibrium points, but it has so-called attracting manifolds of 0-, 1- and 2-dimension. These qualitative techniques for non-stationary phenomena need further study and developments, and, potentially, can be a powerful tool for spatio-temporal electromagnetics [14].

Topological modulation is the discrete change of the spatial content of the field or its "skeletons" according to a control signal. This idea, some processing techniques and circuitry are proposed in [6,8,9,14-20]. Below, we consider a theory of the signal switching mechanism from the point of temporal dynamics of field-force lines [14].

## 2 Passive Gating of Topologically Modulated Signals

The prospective digital hardware for topologically modulated signals should detect the impulses with different time-varying 3-D digital shapes and compare them according to certain logic.

Under the detection it is understood the development of a signal which confirms the arriving of an impulse of a certain topological scheme. Additionally, the topologically modulated signals can be compared with each other or interpreted regarding to a spatial hardware structure similarly to optical holography.

Formally, this hardware needs an established theory of non-autonomous 3-D dynamical systems (1) or (3+1)-D dynamical systems (2) and topological theory of 3-D manifolds and oriented graphs in the (3+1)-D phase space. Unfortunately, these theories are on the development stages. For example, only recently the famous Poincaré conjecture regarding to 3-D manifolds from the 4-D space has been proofed by G. Perelman [12-13].

Our research started in 1988, and, preliminary, the 2-D and 3-D electromagnetic fields had been studied from topological point of view [8]. Then, a qualitative theory of the boundary problems of electromagnetism was created [21]. These techniques allow analytical composition of topological schemes of electromagnetic fields according to the given boundary conditions. The bifurcation parameters for these systems are the frequency or given boundary fields. It was found that the field-force line maps nonlinearly depend on the boundary conditions which smooth variation can change topological schemes of the excited fields.

Later, in 1993, the modal diffraction was considered in a wave transformer from the point of geometry of the diffracted field [16,17]. It was found that the topology of the output signals discretely depends on the input signal magnitudes and their fields. Then, this effect can be used to develop the circuits for processing of topologically modulated signals.

The following example illustrates the effect of passive switching of topologically modulated microwave signals. For this purpose, the circuit that switches the impulses with different topological schemes is represented as a generalized wave transformer (Fig. 1).

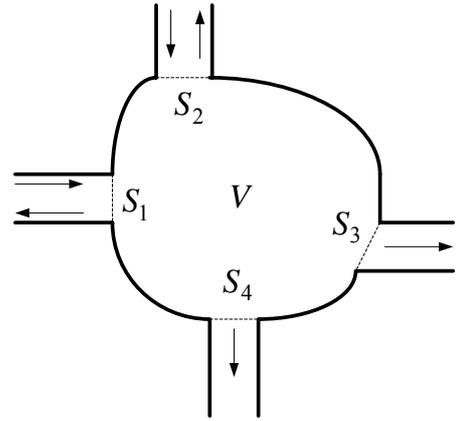

Fig. 2. Wave transformer.

It consists of the central cavity $V$ and $N$ arms. The input fields excite $K$ arms. An example of a four-arm ($N=4$, $K=2$) wave transformer is shown in Fig. 2.

The electromagnetic power conservation law equation is written as:

$$\frac{dW}{dt} = \sum_{k=1}^{K} p_{in}^{(k)} - \sum_{n=1}^{N} p_{sc}^{(n)} \qquad (3)$$

where $W$ is the electromagnetic field energy in the transformer volume $V$, $p_{in}^{(k)}$ is the vector Pointing flow of the incident fields from the $k$-th arm and $p_{sc}^{(n)}$ is the vector Pointing flow of the scattered field to the *n-th* arm.

This expression is written through the field vectors in the central cavity $\mathbf{E}^{(V)}$ and $\mathbf{H}^{(V)}$ and in the *k-th* $(\mathbf{E}_{in}^{(k)}, \mathbf{H}_{in}^{(k)})$ and *n-th* arms $(\mathbf{E}^{(n)}, \mathbf{H}^{(n)})$:

$$\frac{1}{2}\frac{d}{dt}\int_V \left( \begin{array}{c} \varepsilon_a \left(\mathbf{E}^{(V)}(\mathbf{r},t) \cdot \mathbf{E}^{(V)}(\mathbf{r},t)\right)^2 + \\ +\mu_a \left(\mathbf{H}^{(V)}(\mathbf{r},t) \cdot \mathbf{H}^{(V)}(\mathbf{r},t)\right)^2 \end{array} \right) dV =$$
$$= \sum_{k=1}^{K} \int_{s_k} \left[ \mathbf{E}_{in}^{(k)}(\mathbf{r},t) \times \mathbf{H}_{in}^{(k)}(\mathbf{r},t) \right] \mathbf{v}_k ds - \qquad (4)$$
$$- \sum_{n=1}^{N} \int_{s_n} \left[ \mathbf{E}^{(n)}(\mathbf{r},t) \times \mathbf{H}^{(n)}(\mathbf{r},t) \right] \mathbf{v}_n ds.$$

Substituting the electric and magnetic field vectors from (1) the power conservation law (4) is written through the geometrical field characteristics:

$$\frac{1}{2}\frac{d}{dt}\int_V \left( \varepsilon_a \left(\frac{d\mathbf{r}_e^{(V)}(t)}{ds_e}\right)^2 + \mu_a \left(\frac{d\mathbf{r}_h^{(V)}(t)}{ds_h}\right)^2 \right) dV =$$
$$= \sum_{k=1}^{K} \int_{s_k} \left[ \mathbf{E}_{in}^{(k)}(\mathbf{r},t) \times \mathbf{H}_{in}^{(k)}(\mathbf{r},t) \right] \mathbf{v}_k ds - \qquad (5)$$
$$- \sum_{n=1}^{N} \int_{s_n} \left[ \frac{d\mathbf{r}_e^{(n)}(t)}{ds_e} \times \frac{d\mathbf{r}_h^{(n)}(t)}{ds_h} \right] \mathbf{v}_n ds$$

where $\varepsilon_a$ and $\mu_a$ are the absolute permittivity and permeability, respectively, of the media inside the transformer volume and $\mathbf{v}_k$ and $\mathbf{v}_n$ are the normal unit vectors to the opening surfaces of the *k-th* and *n-th* arms, respectively (Fig. 1). This equation shows evolution of the geometry of the excited fields to a steady state if a transient happens with the incident field.

Taking into account that field-force line picture is a particular case of images let's compare the proposed method with the techniques used in signal processing and topology. For example, the images can be enhanced using the differential partial diffusion equation [22]. For a 2-D image an equivalent stationary energy functional is written and the image enhancing is with the minimization of the "image energy." Unfortunately, this method increases the geometrical entropy of the processed image, and it has limitation of its use.

The energy approach was used by G. Perelman for his proof of the Poincare conjecture [12,13]. The used Ricci flow is similar to the nonlinear partial differential diffusion equation. He showed that the time evolution of a closed 3-D manifold of an arbitrary geometry and associated with the Ricci-flow equation leads to the 3-D sphere, and this process is equivalent with reaching the maximum entropy because of the fundamental simplicity of the sphere.

Signal processing is with increase of information [1], and the above mentioned equations have limited applications in logical processing of image-like objects. Here, the time evolution of the geometry should lead to a certain figure which shape not always is tied up with the simplest shape of maximal geometrical entropy.

The equations (3)-(5) derived in 1993 are the energy flow equations, but geometrical solutions of them are not with the maximum of geometrical entropy. The transients in the considered above wave transformer lead to excitation of the cavity and port modes. This field has a certain spatio-temporal geometry which depends on the boundary, initial conditions and transformer geometry. The resonances and non-zero excitations do not allow degrading the field to the noise-like distribution of maximal geometrical entropy. Then, the above considered effects and equations (3)-(5) can be used to describe the information dynamics of such a sort of signal processing devices instead of methods associated with the entropy maximum [11-13,22].

Equations (3)-(5) shows that the geometry of the field-force line maps inside the transformer and in the output arms depends nonlinearly on the input fields and the relationships of their amplitudes. The topological maps can be changed discretely by a smooth variation of the incident field parameters. Particularly, it allows controlling the field distribution of the output fields. If the output field is close to a propagating mode of the *n-th* output, then the matching conditions allow transmitting the maximal power of the incident signal to this output. Other terminals can be isolated from the input if this signal excites the evanescent modes in them. Then, the switching of the input signal from one output to another can be realized choosing the input signal parameters.

Considering the (3+1)-D dynamical systems (2), it follows that it is possible to design the devices handling the signals composed from 3-D manifolds and having their different spatio-time topology [14].

Another conclusion from (3)-(5) is that the discrete dynamics of the vector maps is caused by natural reason of wave diffraction and interference instead of switching mechanisms of semiconductors. Although, the transistor is the best switching device ever made by engineers, the nature realizes digital-like mechanisms using other less-energy consuming means.

## 3 Examples of Passive Gates

The first gates developed to process the topologically modulated signals were of a passive design. They switch the impulses to different outputs according to their spatial "topology". Measurements and simulations were performed in [9,16-20] for digital and microwave signals. An example of such a switch on strip transmission lines is shown in Fig. 2 with its truth-table in Fig. 3.

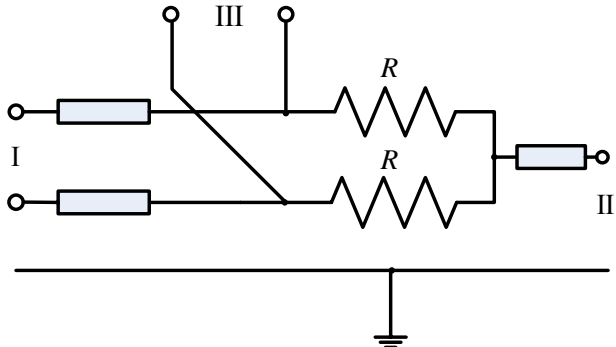

Fig. 2. Passive switch for topologically modulated signals. I- input on the coupled strip line, II- output of the topological "1", and III -output of the topological "0".

The input signal (Port I) is a series of video or sinusoidal impulses of the even and odd modes of coupled strip lines. Their field topological schemes are not homeomorphic to each other, and the even and odd modes correspond to the logical levels "1" and "0", respectively.

| $L \diagdown n$ | **I** | **II** | **III** |
|---|---|---|---|
| "0" | ++× | — | ↔ |
| "1" | ++↕ | ↕ | •• |

Fig. 3. Truth-table for the passive switch of topologically modulated signals realized on strip transmission lines.

The input signals are switched to the outputs II and III according to the truth-table shown in Fig. 3 where $L$ is assigned for the logical levels and $n$ is for the input/output numbers.

The above considered gate switches the signals according their spatial maps only. More enhanced switches can be designed as spatio-temporal filters to process the electromagnetic fields modulated in the (3+1)-D space.

Passive Boolean logic is represented by a microwave OR/AND gate which idealized circuit is shown in Fig. 4 [6,9,23]. It consists of 3 switches $S_{1-3}$ (Fig. 1) of a microstrip design, 3 baluns $B_{1-3}$ and a directional branchline coupler DC. Input signals I and II are microwave impulses of even (logical "1") and odd (logical "0") modes of coupled microstrips. The modes have different topological charts of their fields [20].

The switches analyze the spatial spectrum of incoming impulses, and it corresponds to the first layer of neurons in the brain cortex. The directional coupler DC mixes the spectral content (let's compare with the horizontally oriented neurons). The switch $S_3$ forms the processed image which depends on the spatio-temporal content of incoming impulses and their magnitudes as shown below.

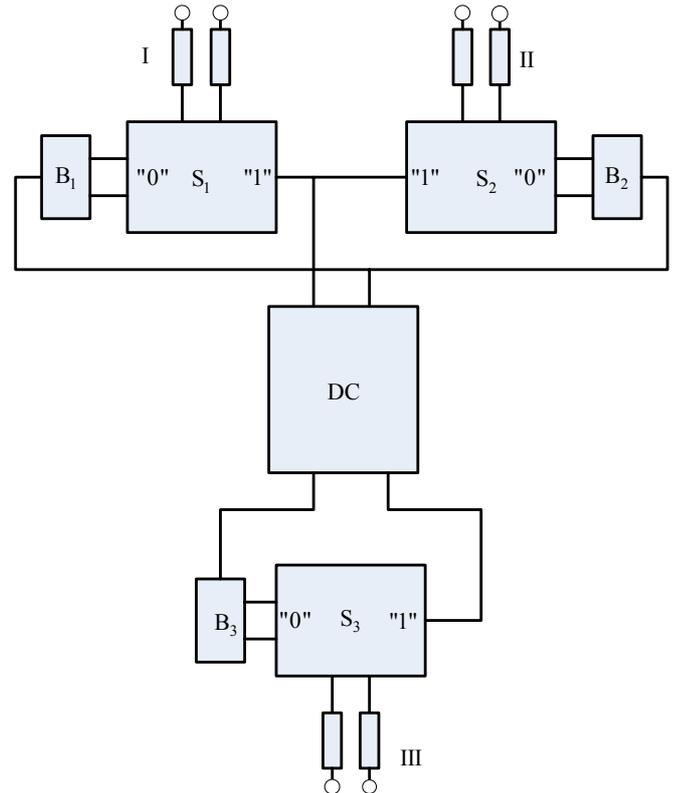

Fig. 4. Passive reconfigurable OR/AND gate for microwave topologically modulated signals.

The results of the express-simulations of this gate were derived by I.V. Nazarov with a SPICE-program, and they are shown in Figs. 5 and 6 for the signals of 10 GHz frequency. The gate was loaded by coupled microstrip lines I-III of the characteristic impedances $Z_e$ =112Ω and $Z_0$ =37.1Ω. The thin-film resistors of a microstrip switch (Fig. 1) are of the value $R$=24.5Ω, and its outputs II and III are calculated for 50Ω and 100Ω loads, respectively. The DC model is an ideal, and it does not take into account its parasitics. The idealized baluns $B_{1-3}$ are to transform the differential and common voltages to each other.

In the case of identical incoming modes, the gate repeats them in the output III. The logical results of different incoming modes are described below.

Fig. 5 is for the incoming even mode (Port I) with impulse magnitude 1 V and the incoming odd mode

(Port II) of the magnitude 0.18 V measured regarding ground. In this case, the curve of the increased magnitude is the voltage of the even mode on the output III, and the gate is working in the regime of logical OR because the even mode corresponds to the logical "1".

Fig. 6 is in the case of the excitation of the input I by low-level even mode (logical "1") with the magnitude 0.18 V. The odd mode (logical "0") voltage is now 1 V measured regarding ground. In this case, this mode suppresses the even one due to the increased magnitude and interference, and the magnitude of the output odd mode is stronger than the magnitude of the even mode. Now, this circuit is working in the regime of the AND gate.

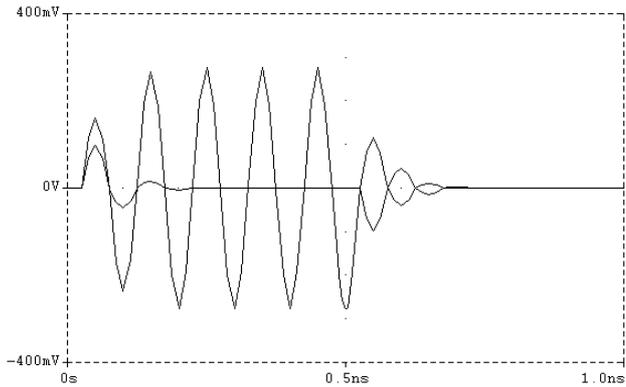
Fig. 5. OR-gate effect. Even mode signal is of the increased magnitude.

The edges of the output impulses are distorted due to transient effects caused by mismatching and parasitics of resistors $R$ in the switches $S_{1-3}$. These effects can be decreased by proper circuit optimization and transition to a monolithic design.

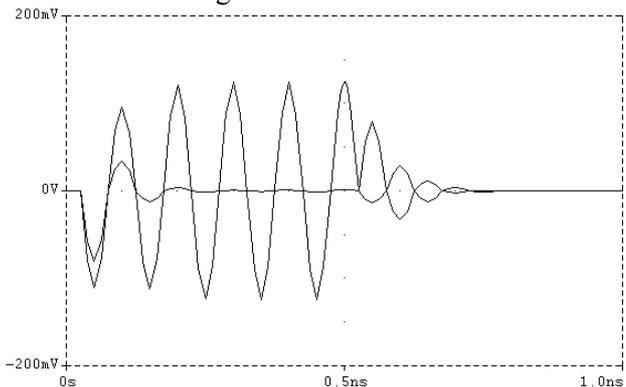
Fig. 6. AND-gate effect. Odd mode signal is of the increased magnitude.

Taking into account the dependence of this logical operation on the magnitude relationship, the above considered circuit OR/AND gate belongs to the class of reconfigurable logic. Additionally, this gate can work as a controlled –NOT gate or a Follower depending on control signal. Composition of larger microwave logic with the above mentioned gates requires more detailed study of degradation of signals due to the transients and loss to define the level of achievable complexity [9]. Other readings on passive logic are from [24-27].

In the brain, signal correction is reached due the active nature of nervous tissue. Electronic technology uses the amplification and gating, and such circuitry for spatio-time signals was developed in [20,28,29]. Similarly to the microwave gates, they analyze the spatial content of video impulses by spatial switches, compare the modal magnitudes and form the output spatial pattern. Some developed gates model the qubit logic and the quasineural pseudo-quantum networks composed of them were proposed. Others model the Boolean or predicate logic or even both and combine the spatial and magnitude logical operations. Recent development is a spatial predicate logic microprocessor which is prospective for data-base management machines [30].

## 4  Acknowledgements
The author thanks I.V. Nazarov (MSIEM-TU, Moscow, Russia) for his help in simulation of circuits.

## 5  Conclusion
It has been supposed that the digital-like activity of the brain, which is mostly a quasi-linear and active system, can be caused, partly, by non-topological transformations of spatio-temporal patterns in the brain cortex. The topology of a pattern corresponds to a logical unit, and the elementary logical operations occur even during superposition of spatio-temporal signals in such a linear system as the brain. Patterned signals have been modeled by spatio-temporal topologically modulated impulses of the electromagnetic field, and their theory and switching mechanism of them have been proposed. A particular case of a reconfigurable gate OR/AND for microwave signals and its simulation results have been considered. For digital signals, a short review on spatial logic hardware has been given. The derived results are interesting for modeling of neural networks and the development of new computing hardware.